\documentstyle[prl,aps,multicol]{revtex}
\input psfig
\def \be{\begin{equation}}
\def \ee{\end{equation}}

\def \o{\omega}

\def \e{\epsilon}

\def \H{{\cal H}}
\def \K{G^{(K)}}
\def \R{G^{(R)}}
\def \A{G^{(A)}}

\def \D{{\cal D}}

\def \varQ{{\overline{Q^2}}}
\def \mls{\delta_1}
\begin{document}

\bibliographystyle{simpl1}

\title{Charge Pumping and Photovoltaic effect in Open Quantum Dots.} 
\date{\today}
\author{Maxim~G.~Vavilov and V.~Ambegaokar}
\address{Laboratory of Atomic and Solid State Physics,
Cornell University, Ithaca, NY 14853}
\author{Igor L.~Aleiner}
\address{
Department of Physics and Astronomy, SUNY at Stony Brook,
Stony Brook, NY 11794
}
\maketitle
\begin{abstract}
We propose a random matrix theory to describe the influence of a 
time-dependent
external field on electron transport through open quantum dots.
We describe the generation of the current by an oscillating field for the
dot, connected to two leads with equal chemical potentials.
For low frequency fields our results correspond to adiabatic
charge pumping. Finite current can be produced if the system goes
along a closed loop in parameter space, which covers a finite
area.
At high frequency a finite current is produced even if the loop is
a line in parameter space. This result can be explained in the same
way as adiabatic pumping but considering the evolution of the system
in phase 
space rather than in parametric space.
\end{abstract}
\draft
\begin{multicols}{2}

\section{Introduction}
Adiabatic charge pumping through open quantum dots 
was studied recently in the literature both theoretically \cite{Br,SZA,SAA}
and experimentally \cite{M&Co}. Such pumping occurs in a
system described by a Hamiltonian periodic in time with a
period $T_{\rm p}$ larger than all other characteristic time scales of
the system.  
After one period, the system returns to its initial form; however  
charge $Q$ can be transmitted through a cross--section of the system:
\begin{equation}
\label{0}
Q=I_{\rm DC}T_{\rm p}=\int_0^{T_{\rm p}}\langle I(t)\rangle dt,
\end{equation} 
where $\langle \dots \rangle$ denotes quantum mechanical and
thermodynamic averaging.

To obtain a finite transmitted charge at low frequencies, the
Hamiltonian should depend on at least two parameters. 
In Refs.\cite{Br,SZA,SAA} the time dependence of the Hamiltonian was replaced
by a dependence on parameters and the system was considered
quasistationary for each parameter value. The transported charge 
during one period of the Hamiltonian was calculated as an integral in 
the parameter space. The theory \cite{SAA} shed some light on the recent 
experiments \cite{M&Co}, namely on the amplitude dependence of the
root mean square fluctuations of the transmitted charge, averaged
over different realizations of the Hamiltonian. 

A very similar phenomenon was considered previously by Falko and
Khmelnitskii, who theoretically studied  
the photovoltaic effect in mesoscopic microjunctions \cite{FK}. 
The experimental observation is described in Ref. \cite{Kvon}. The
photovoltaic 
effect is a generation of d.c.-current by radiation of a finite
frequency. (It is 
obvious that this effect can only be non--linear in the oscillating
field.) The bilinear  
regime of adiabatic pumping, \cite{Br,SZA,SAA} is precisely the circular
photovoltaic effect introduced in Ref.~\cite{BS} and applied to
a mesoscopic system in Ref.~\cite{FK}. The results of
Ref.~\cite{FK} are not 
directly applicable to quantum dots because in microjunctions the
Thouless energy
$E_{\rm T}\sim 1/\tau_{\rm erg}$
is of the same order as the inverse escape time $1/\tau_{\rm esc}$,
whereas for 
quantum dots $1/\tau_{\rm esc}\ll E_{\rm T}$. (Here $\tau_{\rm
erg}$ is the characteristic time for a classical 
particle to cover all of the available phase space in the dot and
we put $\hbar=1$.) 
Therefore the considerations of
Refs. \cite{Br,SAA} 
have their own physical significance. On the other hand, the theory of
Ref. \cite{FK} 
is not restricted to the adiabatic regime, the results being valued in 
a broad interval of frequencies.

The purpose of the present work is to go beyond the adiabatic
approximation for d.c.-current generation in open quantum dots.  
One can identify two contributions to the d.c.-current --- reversible
and irreversible. To
make a connection with the terminology of the photovoltaic effect, used in
Ref.~\cite{FK}, we consider the bilinear d.c.-current response
through the dot, generated by several time-dependent perturbations
$\varphi_i(t)=\varphi_{i,\omega}e^{i\omega t}$. 

The direct current $I_{\rm DC}$ can be written at $\omega\to 0$ as
\begin{equation}
\label{2linear}
I_{\rm DC}=i\omega
\epsilon_{ij}\varphi_{i,\omega}\varphi^*_{j,\omega}+
\omega^2\gamma_{ij}\varphi_{i,\omega}\varphi^*_{j,\omega}+{\cal O}(\omega^3),
\end{equation} 
where $\epsilon_{ij}$ and $\gamma_{ij}$ are real sample specific
coefficients. They are not fixed by any symmetry (a sample does not
have any!) except the condition that $I_{\rm DC}$ is a real number,
which gives the requirements
\begin{equation}
\label{eg}
\begin{array}{cc}
\epsilon_{ij}=-\epsilon_{ji}, & \gamma_{ij}=\gamma_{ji}.
\end{array}
\end{equation}
These relations mean that as the direction of the contour in the
parameter space 
$\{ \varphi_i(t) \} $ is reversed ($\varphi_i(t)\to \varphi_i(-t)$ or 
$\varphi_{i,\omega} \to \varphi^*_{i,\omega}$), the first term changes
its sign whereas the second term remains intact. In the language of
the photovoltaic effect, the first term is the circular photovoltaic
effect, and the second term is the linear effect. \cite{FK} Equation
(\ref{2linear}) makes an explicit 
connection between the bilinear responses of d.c.-current generation 
through open quantum dots and the photovoltaic effect.

We will see that the separation onto reversible and irreversible parts
goes far beyond the bilinear and low frequency expansions. We will find
that the reversible part vanishes at high frequency whereas the
irreversible part saturates. We will also show that the irreversible
contribution may be interpreted in a manner similar to adiabatic pumping. In 
adiabatic pumping the transmitted charge was determined by a
contour in parameter space. The irreversible current is
determined by the contour in the extended phase space $\{\varphi_i(t),
\dot \varphi_i(t), \dots \}$, see Sec.~V.

The first term in Eq.~(\ref{2linear}) vanishes for a single 
pump due to the antisymmetry of
$\epsilon_{ij}$, Eq.~(\ref{eg}).
The contour in parameter space
degenerates to a line in this case, while in phase space
the contour encompasses a finite area. We will see that the current is
proportional to this area. Note, that the contour is invariant with
respect to time inversion.

The remainder of the paper is organized as follows.
In section II we formulate the model which is studied in the paper.
We mainly discuss the high temperature limit $T\gg\omega$, where $\omega$
is the frequency of the perturbation. In section III we calculate the
ensemble averaged fluctuations of the current. As an example, 
low frequency asymptotic of reversible (adiabatic) and irreversible
currents are considered. In Sec. IV we discuss how the irreversible
contribution can be represented in the form of a linear integral over
the contour in extended phase space. In Sec. V we discuss the low
temperature limit and show how to take heating into account.
Section VI summarizes our results.


\section{The model}


We consider the following experimental realization of the model.
Gates near a two--dimensional electron gas (2DEG) form the shape of
the dot. An oscillating 
voltage is applied to the gates $V_1(t)$ and $V_2(t)$. As a result of
motion of electron energy levels in the dot, 
a current flows through the dot. The direction of the current depends
on the particular realization of the dot and is zero on average.
We calculate the fluctuations of the d.c.-current with respect to
different realizations of the dot. 

Calculations will be performed for an open quantum dot 
in the limit of a large number of open channels $N_{\rm ch}$
connecting the dot to the leads. 
This condition allows us to neglect the
electron--electron interaction, which gives corrections of the
$1/N_{\rm ch}^2$ order, (See Ref. \cite{Brouwer}). The same condition 
permits the use of a diagrammatic technique, similar to that described in
\cite{AGD}, to calculate ensemble averaging.

We also assume that the quantum dot 
is small and the Thouless energy $E_{\rm T}\sim 1/\tau_{\rm erg}$ is much
greater than all other energy scales of the problem.
In this limit, one can use random matrix theory (RMT) to 
study transport and thermodynamic properties of the system, see
Ref. \cite{B}. All corrections to the RMT are as small as
$N_{\rm ch}/g_{\rm dot}$, where $g_{\rm dot}=E_{\rm T}/\mls$, 
$\mls$ is the mean level spacing. 
We neglect the fluctuations of the time dependent perturbation from
sample to sample, created by the voltages
$V_{1,2}(t)$, since they depend on the small parameter
$g^{-1}_{\rm dot}\ll 1$.

\begin{figure}
\centerline{\psfig{figure=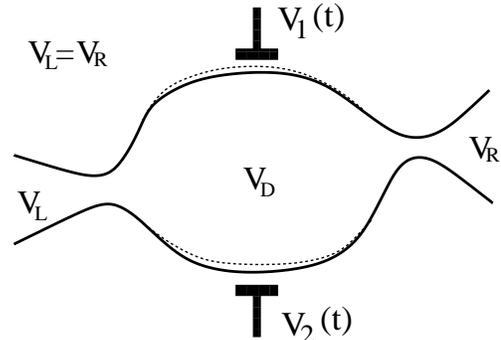,width=6.5cm}}
\narrowtext{
\caption{
An experimental setup. The voltage applied to the gates $V_1(t)$ and $V_2(t)$
changes the shape of the dot, resulting in motion of the energy levels 
of the electrons in the dot.}}
\end{figure}

The Hamiltonian of the system is
\cite{B}
\be
\label{1}
\hat H=\hat H_{\rm D}+\hat H_{\rm L}+\hat H_{\rm LD},
\ee
where $\hat H_{\rm D}$ is the Hamiltonian of the electrons in the dot,
which is determined by the $M\times M$ matrix $H_{nm}$:
\be
\label{2}
\hat H_{\rm D}=\sum_{n,m=1}^M \psi^\dag_n H_{nm} \psi_m,
\ee
where the thermodynamic limit $M \to\infty$ is assumed.

The coupling between the dot and the leads is
\be
\label{7}
\hat H_{\rm LD}=\sum_{\alpha, n, k}\left( W_{n \alpha}\psi^\dag_\alpha
(k)
\psi_n+
{\rm H.c.}\right),
\ee
where $\psi_n$ corresponds to en electron state in the dot,
$\psi_{\alpha}(k)$ denotes electron state $(\alpha, \ k)$  in the leads,
and $k$ labels the continuum of momentum states in each channel $\alpha$.
For a dot connected by two leads with $N_{\rm l}$ and
$N_{\rm r}$ channels respectively,
we denote the left lead channels by
$1\leq \alpha\leq N_{\rm l}$ and the right channels by
$N_{\rm l}+1\leq \alpha \leq N_{\rm ch}$, where
$N_{\rm ch}=N_{\rm l}+N_{\rm r}$.
The coupling constants, $W_{n \alpha}$,
in Eq.(\ref{7}) are \cite{B}:
\be
\label{9}
W_{n \alpha}=\cases{
\Gamma_n,& if $n=\alpha\leq N_{\rm ch}$,\cr
0,& otherwise
}
\ee
and $\Gamma_n$ are defined below in Eq.~(\ref{16z})
 
The electron spectrum in the leads near Fermi
surface can be linearized
\be
\label{8}
\hat H_{\rm L}=v_{\rm F}\sum_{\alpha, k} k \psi^\dag_\alpha(k)\psi_\alpha(k),
\ee
where  $v_{\rm F}=1/2\pi \nu$ is the
Fermi velocity, and $\nu$ is the density of states per channel at the
Fermi surface.

The current through the dot is given in terms of the scattering
matrices $\hat {\cal S}(t,t')$ by the following expression,
see appendix A:
\begin{eqnarray}
\label{10}
\langle I(t) \rangle & = & 
e \sum\limits_{\alpha}\Lambda_{\alpha\alpha}
\int\! dt_1dt_2
\\
\displaystyle
&\times &
\left\{
\sum\limits_{\beta}
{\cal S}_{\alpha\beta}(t,t_1)f_\beta(t_1-t_2)
{\cal S}^{\dagger}_{\beta\alpha}(t_2,t)
-f_\alpha(+i0)
\right\}
,
\nonumber
\end{eqnarray}
where $f_\alpha(t)$ is the Fourier transform of the electron distribution
function in the leads, $\langle \dots \rangle$ stands for the quantum
mechanical and thermodynamic averages for a given ensemble
realization (no ensemble averaging!) and 
\be
\label{11}
\Lambda_{\alpha\beta}=\delta_{\alpha\beta}\cases{\displaystyle
+\frac{N_{\rm r}}{N_{\rm ch}}, & if $1\leq \alpha \leq N_{\rm l}$;\cr
\displaystyle 
-\frac{N_{\rm l}}{N_{\rm ch}}, & if $N_{\rm l}< \alpha \leq 
N_{\rm ch}$.\cr}
\ee
The scattering matrix of the system,
$\hat {\cal S}(t,t')$, is
\be
\label{13}
{\cal S}_{\alpha\beta}(t,t')=\delta_{\alpha\beta}\delta(t-t')-2\pi i\nu
W^\dag_{\alpha n} \R_{nm}(t,t') W_{m\beta},
\ee
and the Green's function $\R_{nm}(t,t')$ is the solution to:
\be
\label{14}
\left(i\frac{\partial}{\partial t}-{\hat H}(t)+
i\pi\nu \hat{W}\hat{W}^\dag \right)\hat G^{(R)}(t,t')=
\delta(t-t'),
\ee
where the matrices $\hat{H}$ and $\hat{W}$ are
comprised by their elements (see Eqs.~(\ref{2}) and (\ref{9})).
An analogue of Eq.(\ref{10}) was used before in the energy representation 
in Ref. \cite{BTP}.

Below we consider the special case of electrons in both leads being 
described by identical distribution functions $f_j(t)\equiv f(t)$.
The function $f(t)$ is the Fourier transform of the Fermi--Dirac
distribution
function:
\be
\label{12}
f(t)=\int\limits_{-\infty}^{+\infty}\!\frac{d\omega}{2\pi}\ \  e^{i\omega t}
\left\{ \frac{1}{e^{\omega/T}+1}-\frac{1}{2}\right\}=
\frac{i T }{2\sinh\pi T t}
\ee

In this case, we can derive (see Appendix B)
 another formula for the current through the dot:
\begin{eqnarray}
\label{19}
\displaystyle
\langle I(t)\rangle 
& = & 2e i \pi \nu {\rm Tr} \int\!\!\int\!\!dt_1dt_2 f(t_1-t_2)
\\
\displaystyle
& \times &
\left\{
\hat W^\dagger \hat \R (t,t_1)\left[\hat H(t_1)-\hat H
(t_2)\right]\hat \A(t_2,t)\hat W\Lambda\right\},
\nonumber
\end{eqnarray}
which is more convenient for further calculations.

We calculate the variance (mean square) of the transported charge $Q$
through the dot. 
We assume, that the Hamiltonian of the dot $\hat H$ in Eq.(\ref{2}) is
represented by a time--dependent matrix in the form:
\be
\label{3}
\hat H(t)=\hat \H+\hat V_1 \varphi(t) +\hat V_2 \psi(t).
\ee
Here, the time independent part of the Hamiltonian $\hat \H$ is
a random $M\times M$ matrix, which obeys Gaussian statistics with
the correlator
\be
\label{4}
\overline{\H_{nm}\H^*_{n'm'}}=\lambda\delta_{nn'}\delta_{mm'}
+\lambda'\delta_{mn'}\delta_{nm'}, 
\ee 
where $\overline{(\dots)}$ means averaging over ensemble of matrices
$\hat \H$, 
$\lambda=M(\delta_1/\pi)^2$ and $\lambda'=\lambda(1-g_{\rm h}/4M)$,
and $g_{\rm h}$ defines the crossover from
orthogonal ($g_{\rm h}=0$) to unitary ($g_{\rm h}=4M$) ensembles.
The parameter $g_{\rm h}$
has the meaning of the dephasing rate due to an external magnetic field in
units of the level spacing $\delta_1$, \cite{B,AAKL}.
It can be estimated as $g_{\rm h} \simeq g_{\rm dot} 
\left(\Phi/\Phi_0\right)^2$
where $\Phi$ is the magnetic flux through the dot and $\Phi_0=hc/e$ is
the flux quantum. The time dependent perturbation is described by
symmetric $M\times M$ matrices, $\hat V_i$, and the time dependent
functions $\varphi(t)$ and $\psi(t)$.

We make the simplifying assumption that the matrices $\hat V$ are
traceless ${\rm Tr}\hat V=0$.
In this case, the unknown parameters $C_{ij}$ are defined as
\be
\label{5}
{C}_{ij}=\frac{\pi }{M^2 \mls} {\rm Tr}\hat V_i\hat V_j,
\ee
where we used the fact that the matrix $\hat{V}$ is symmetric.  
(We note that according to the definition of Eq.(\ref{5}), $\hat C$
has dimension of energy.)
The parameters $\hat C$ are also related to the typical value of the level
velocities which characterize the evolution of  an energy level
$\epsilon_{\nu}({\bf X})$ under
the external perturbation ${\bf X}\hat{{\bf V}}$, \cite{E}:
\be
\label{6}
\frac{2\mls}{\pi} C_{ij}=
\overline{
\frac{\partial \e_\nu}{\partial X_i}
\frac{\partial \e_\nu}{\partial X_j}
}
-
\overline{\frac{\partial \e_\nu}{\partial X_i}}
\
\overline{\frac{\partial \e_\nu}{\partial X_j}}.
\ee

Since all other responses (e.g., parametric dependence of the
conductance of the dot) are expressed in terms of 
universal functions of the same parameters $C_{ij}$ \cite{E}, they can
be found from independent measurements.
Calculations can be generalized to the case in which the trace of the
perturbation matrix is not zero. See, e.g. \cite{SAA}.

For reflectionless contacts, the ensemble averaged scattering matrix 
$\overline{{\cal S}_{\alpha\beta}} $
is zero. This condition determines the factors $\Gamma_n$ in
Eq.~(\ref{9}):
\begin{equation}
\label{16z}
\Gamma_n= \sqrt{\frac{M\delta_1}{\pi^2\nu}}.
\end{equation}

The ensemble average of the transmitted charge $Q$, defined by
Eq.(\ref{0}), is zero: $\overline{Q} = 0$. We calculate the second
order correlator with respect to an ensemble of random matrices,
Eq.~(\ref{4}):  
\be
\label{21}
\overline{Q^2} =\int\limits_0^{T_{\rm p}}
\!\int\limits_0^{T_{\rm p}}\! dtdt'
\overline{\langle I(t)\rangle\langle I(t')}\rangle.
\ee
For this purpose we use a diagrammatic technique, which has been applied to
a similar class of problems. See Refs.\cite{SAA,VA}.
Here we present the basic elements of the diagrams which will appear
in sections III and V.
The ensemble averaged Green's function $G_{nm}^{R,A}$ in the absence
of perturbations is equal to
\be
\label{15}
\overline{G_{nm}^{R,A}(\e)}=\pm\frac{\delta_{mn}}{i \sqrt{\lambda M}}
\cases{\displaystyle
 1+\frac{N_{\rm ch}\pm i \e}{4M} , &
$N_{\rm ch}<n \leq M$; \cr
\displaystyle
{1 \over 2}, & $1\leq n \leq N_{\rm ch}$.
}
\ee
Above we introduced the dimensionless energy, $\e$, measured in units of
$\sqrt{\lambda/4M}=\delta_1/2\pi$. We expanded these Green's
functions in $\e/M$ and $N_{\rm ch}/M$, since only those terms survive
the thermodynamic limit
$M\to\infty$. For the same reason, in the expression for $\R_{n}$ with
$n\leq N_{\rm ch}$  one has to neglect such terms since the contribution of
these elements to the final result is already of the order of $N_{\rm ch}/M$.

The other element of the diagram technique used in the 
paper is an amputated average of the
products of two Green's functions (see Fig.2(b)), 
called the diffuson $\D(t_1,t_2,\tau)$ and defined by 
\begin{eqnarray}
& & 
\overline{{\big (}\R_{nm}(t_1^+,t_2^+)
\A_{mn}(t_2^-,t_1^-){\big )}}_{\rm amp} 
\label{16a}
\\
& = & 4M\lambda
\D\left(\frac{t_1^++t_1^-}{2},\frac{t_2^++t_2^-}{2},t_1^+-t_2^+\right)
\nonumber
\\
& \times &
\delta(t_1^+-t_2^+-t_1^-+t_2^-)  
\nonumber
\end{eqnarray}
We can use this relation since
the time arguments of diffuson satisfy
$ t_1^+-t_2^+=t_1^--t_2^-$. Introducing new variables
$t_{1,2}=(t^+_{1,2}+t^-_{1,2})/2$ and $\tau=t^+_{1,2}-t^-_{1,2}$
we obtain the following equation for the diffuson $\D(t_1,t_2,\tau )$:
\begin{eqnarray}
\displaystyle
\left[\frac{\partial }{\partial t_1}  +  
{\cal K}_{\D}(t_1,\tau)\right]
{\cal D}(t_1,t_2,\tau)=\delta(t_1-t_2),
\nonumber
\\
\label{17}
\displaystyle
{\cal K}_{\D}(t,\tau)  = N_{\rm ch}+ \Phi^T(t,\tau) \hat C \Phi (t,\tau),
\end{eqnarray}
where
\begin{equation}
\label{18a}
\Phi(t,\tau)=\left(
\begin{array}{c}
\phi \left(t+\frac{\tau}{2}\right)-\phi \left(t-\frac{\tau}{2}\right)
\\ 
\psi \left(t+\frac{\tau}{2}\right)-\psi \left(t-\frac{\tau}{2}\right)
\end{array}
\right).
\end{equation}
The solution to the above equation is
\be
\label{18}
\D(t_1,t_2,\tau)=\Theta(t_1-t_2)\exp\left(-\int\limits_{t_2}^{t_1}
{\cal K}_{\D}(t,\tau )dt  \right).
\ee
Equations (\ref{17})--(\ref{18}) are written in dimensionless variables, so
that energy and time are measured in units of $\mls/2\pi$ and $2\pi/\mls$
respectively. Below, intermediate expressions are also written in
terms of dimensionless energy and time variables, while the final answers are
represented in terms of physical quantities.

\begin{figure}
\centerline{\psfig{figure=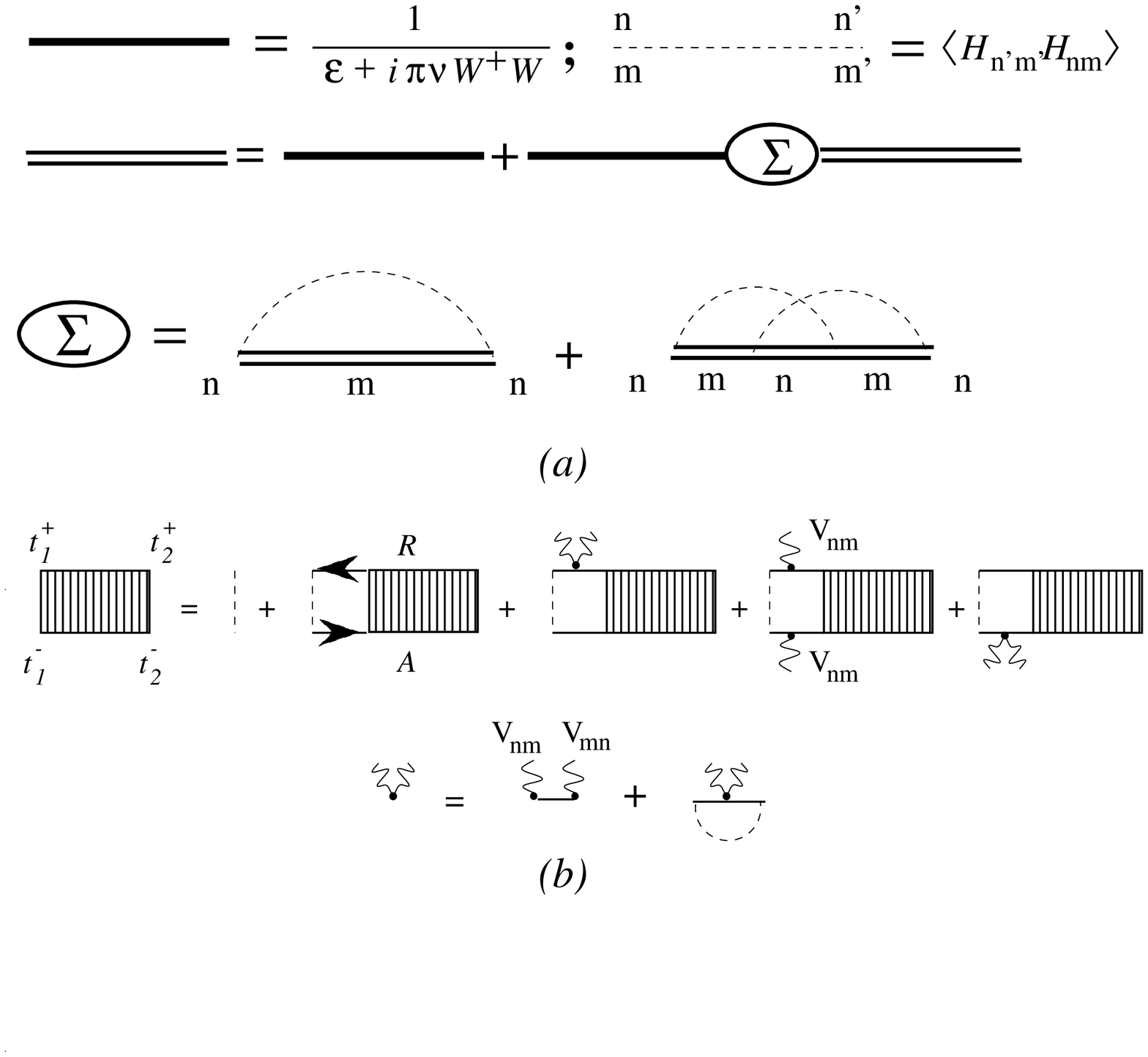,width=8cm}}
\narrowtext{
\caption{
(a) Diagrams for the ensemble averaged Green's function.
The second term in the self-energy
includes an intersection of dashed lines and is as
small as $1/M$. (b) The Dyson type equation for the diffuson, 
${\cal D}(t_1^+,t_1^-,t_2^+,t_2^-)$.
}
}
\end{figure}

\begin{figure}
\centerline{\psfig{figure=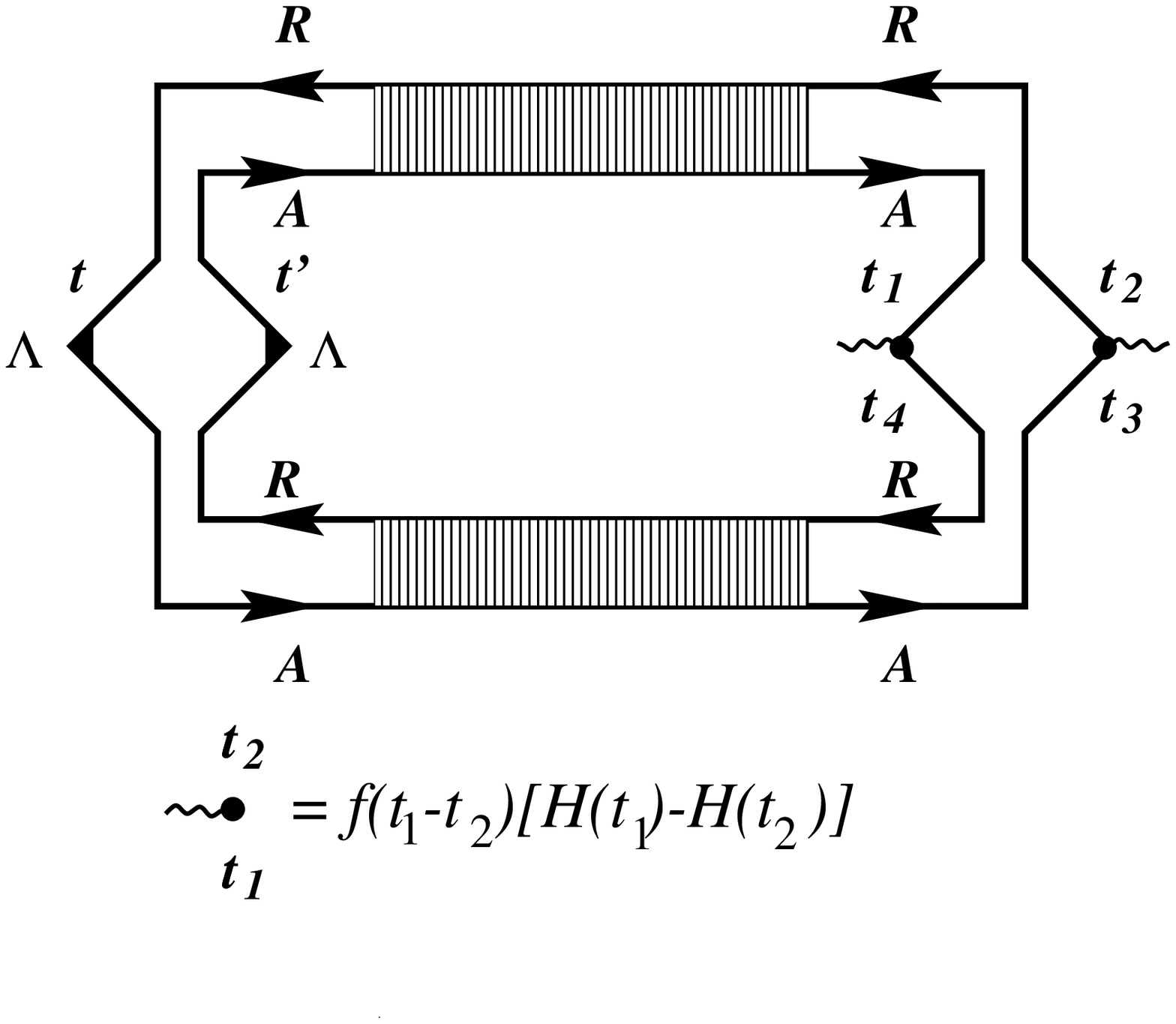,width=7.5cm}}
\narrowtext{
\caption{
The diagram representing the contribution to the charge correlator,
$\varQ$ at high temperature $T$.}} 
\end{figure}
  

\section{Mesoscopic fluctuations of the Current at high temperatures}


In this section we will consider the high temperature limit, in which the
frequency of the perturbation is much smaller than the temperature.
(More accurate definition of the high temperature limit is given in Sec.~V.)
In this case the only contribution to the charge correlation function,
$\varQ$, is given by the diagram shown in Fig.~3.
The corresponding analytical expression is
\begin{eqnarray}
\displaystyle
\varQ&=& 4 e^2 g
\int\limits_0^{2\pi}\!\! dxdy {\cal R}(x,y)
\int\limits_0^{+\infty}\!\!d\theta \int\limits_{-\theta}^{\theta}\!\! d\tau
F^2(\tau)
\nonumber
\\
\displaystyle
\label{22}
&\times &
\D\left(\frac{x+y}{2\omega}+\theta,
\frac{x+y}{2\omega}+\tau,\frac{x-y}{\omega}\right)
\\
&\times &
\D\left(\frac{x+y}{2\omega}+\theta,
\frac{x+y}{2\omega}-\tau,\frac{x-y}{\omega}\right),
\nonumber
\end{eqnarray}
where
\begin{eqnarray}
\displaystyle
{\cal R}(x,y)&=&C_{11}\sin x\sin y+
C_{22}\sin(x+\phi)\sin(y+\phi)
\nonumber
\\
& +& C_{12}\left(\sin(x+\phi)\sin y+\sin x\sin(y+\phi)\right),
\label{23}
\end{eqnarray}
\begin{equation}
\label{g}
g=\frac{N_{\rm l}N_{\rm r}}{N_{\rm ch}}
\end{equation}
is the dimensionless 
conductance through the dot from the left to right leads and
\begin{equation}
\label{223}
F(\tau)=\frac{ T \tau}{\sinh 2 \pi T \tau}.
\end{equation}
Expression (\ref{22}) can be computed for different
values of parameters.
In Fig.~4 we present the result of computation of $\varQ$ 
for two frequencies $\omega=0.1\gamma_{\rm esc}$ and 
$\omega=\gamma_{\rm esc}$. Both those curves exhibit $C_{\rm l}^2$ and
$\sqrt{C_{\rm l}}$ dependences 
at weak and strong pumping respectively. We also show the
analytical curve given by Eq.(\ref{33}) for the $\omega\to \infty$ limit.
Below we discuss different limits of Eq.(\ref{22})

\begin{figure}
\centerline{\psfig{figure=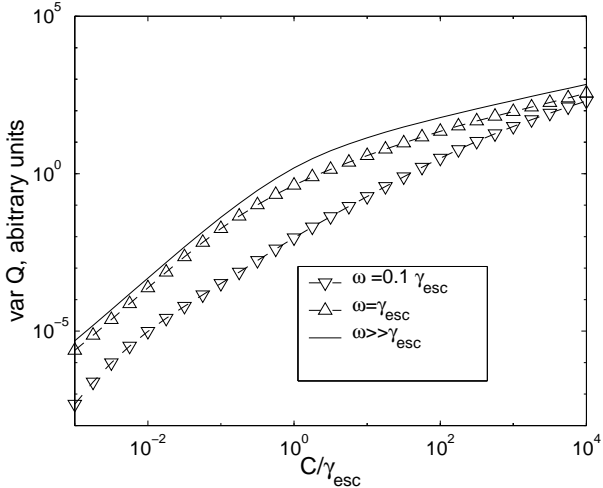,width=8.1 cm}}
\label{fig4}
\narrowtext{
\caption{ 
The amplitude dependence of the pumped charge for different values of 
the frequency of the pump. For $\omega \geq \gamma_{\rm esc}$ the curves
have the
$C^2$ dependence at small values of $C$ and the $\sqrt{C}$ dependence
at $C\gg \gamma_{\rm esc}$. For small frequency (e.g.
$\omega=0.1\gamma_{\rm esc}$) there
is an intermediate regime.  
}}    
\end{figure}


\subsection{Bilinear response}


First we consider the weak perturbation and perform an 
expansion of the diffusons up to terms linear in $C_{ij}$.
As a result we obtain:
\begin{eqnarray}
\label{27}
\displaystyle
\varQ & = & 4\pi^2 e^2 g \int_0^{\infty}\!\!d\theta e^{-2N_{\rm ch}\theta}
\int_{-\theta}^{+\theta} d\tau F^2(\tau)
\nonumber
\\
\label{28}
\displaystyle
&\times & 
\frac{C_{\rm l}^2 (2\omega \theta -\sin 2\omega \theta )
+ C_{\rm c}^2 \sin 2\omega \theta }{\omega},
\end{eqnarray}
where we have introduced the linear and circular pumping amplitudes
\begin{eqnarray}
\displaystyle
\label{29}
C_{\rm l} & = & C_{11}+2C_{12}\cos \phi+C_{22},\\
\displaystyle
\label{29.1}
C_{\rm c} & = & 2 \sin\phi \sqrt{C_{11}C_{22}-C_{12}^2}.
\end{eqnarray}
In the case of temperature $T$ larger than the escape rate
$\gamma_{\rm esc}=N_{\rm ch}\mls/2\pi$, ($T\gg \gamma_{\rm esc}$) we find
\begin{eqnarray}
\label{30}
\displaystyle
\varQ = \frac{e^2}{24} g \frac{\mls}{T}
\frac{1}{\gamma_{\rm esc}^2+\omega^2}
\left(
\frac{\omega^2}{\gamma^2_{\rm esc}}C^2_{\rm l}
 + C_{\rm c}^2
\right).
\end{eqnarray}

The second term of Eq.(\ref{30}) survives the limit $\omega\to 0$, thus
reproducing the known result for adiabatic pumping \cite{Br,SAA}.
On the other hand, this term vanishes at high frequency. 
The linear term is quadratic in frequency at small frequency and 
tends to a constant at large frequency. 

The linear pumping amplitude $C_{\rm l}$ in the case of two pumps has the
form of Eq.(\ref{29}), which implies that the amplitude is just a
vector sum of different pumps in the parameter space. On the other
hand the circular amplitude is related to the area in the parameter
space, covered by the pumps.  
 

\subsection{Low frequencies}


Equation~(\ref{22}) in the adiabatic limit $\omega\to 0$ is in agreement
with the results of Ref.\cite{SAA}. Namely, this expression gives  
the same asymptotic behavior for the limits of weak and strong pumping.
To demonstrate this we consider the special case of the $\hat C$ matrix
having the form $C_{11}=C_{22}=C$ and $C_{12}=0$.
In this case we obtain 
\begin{eqnarray}
\label{25}
\displaystyle
\varQ&=& \frac{1}{12} e^2 g \frac{\mls}{T}
\frac{2C+(\gamma_{\rm esc}-
\sqrt{\gamma_{\rm esc}(\gamma_{\rm esc}+4C)})}
{\sqrt{\gamma_{\rm esc}(\gamma_{\rm esc}+4C)}}.
\end{eqnarray}
As temperature drops down to $T\sim \gamma_{\rm esc}=N_{\rm ch}\mls/2\pi$, 
the variance of the
transmitted charge saturates to
\begin{equation}
\label{25lt}
\varQ=\frac{2}{\pi}e^2g \frac{\mls}{\gamma_{\rm esc}}
\frac{C^2}{\sqrt{\gamma_{\rm esc}(\gamma_{\rm esc}+4C)^3}}
\end{equation}

The authors of Ref.\cite{SAA} showed that for strong pumping
the mean square transported charge is proportional to the length
of the contour in 
the parameter  space, and does not depend on the particular shape of
the contour. Equations~(\ref{25}) and (\ref{25lt}) support this statement,
since for $C\gg \gamma_{\rm esc}$ they reproduce a
$\sqrt{C}$ dependence on the pumping amplitude.  In the opposite case
of weak pumping Eqs.~(\ref{25}) and (\ref{25lt}) give $C^2$ dependence in
accordance with \cite{Br,SZA}.
To understand the strong pumping dependence, we consider a loop in
the parameter space. See Fig.(5). We notice (see Eq.(\ref{12c1})) and
Refs.~\cite{Br,SAA}, that adiabatic pumping can be related to a
contour integral in the parameter space. At sufficiently strong pumping the
system at distant points of this space is uncorrelated and the total
contribution to the pumped charge comes from the uncorrelated pieces
of the loop, being proportional to their number.

\begin{figure}
\centerline{\psfig{figure=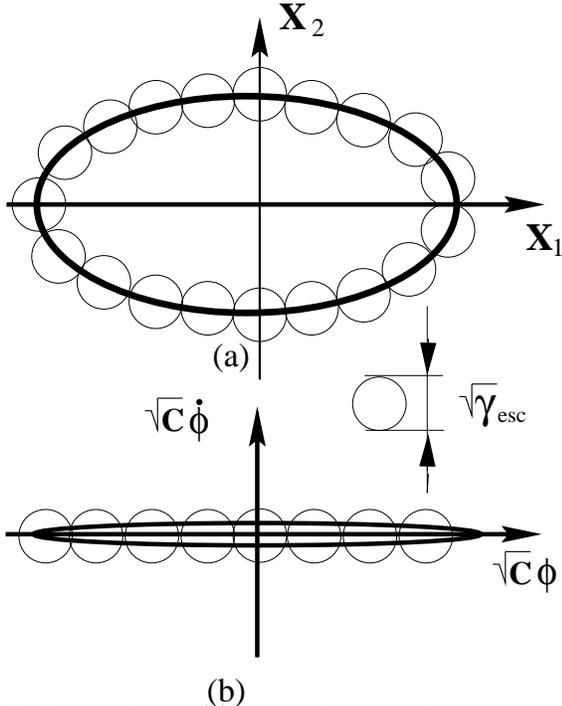,width=7.5cm}}
\label{fig5}
\narrowtext{
\caption{
In figure (a) a loop is shown in the parameter plane. The grid
divides the plane onto pieces, so that parts of the loop
in the different pieces give uncorrelated contributions to the
transported charge. Figure (b) shows the loop in the phase space for 
strong pumping. In this case the loop can be divided onto pairs, and
the pairs are not correlated. On the other hand parts of the loop
of one pair are close to each other, so they are strongly correlated.
}} 
\end{figure}

In the limit of low frequency and zero $C_{\rm c}$ (single pump), 
the mean square fluctuation of the charge per cycle
is quadratic in frequency.  
For weak pumping the amplitude of charge
fluctuations is determined by Eq.(\ref{30}) with $\phi =0$ for
arbitrary frequency $\omega$. (For a single pump, $C_{\rm l}$ is the
only parameter.) 
For strong pumping (but still 
$\omega^2C_{\rm l}\ll \gamma_{\rm esc}^3$) we find
\begin{equation}
\label{32}
\varQ = \frac{25}{576\pi} e^2 g \frac{\omega^2}{\gamma_{\rm esc}^2}
\frac{\mls}{T}\left(\frac{C_{\rm l}}{\gamma_{\rm esc}}\right)^{3/2}.
\end{equation}
We explain this dependence on the
amplitude of the perturbation in the next section.


\subsection{High frequencies}


In the limit of high frequencies, $T\gg \omega\gg \gamma_{\rm esc}$, 
the variance of the transmitted charge is given by
\begin{eqnarray}
\label{33}
\displaystyle
\varQ&=&\frac{1}{12} e^2 g \frac{\mls}{T}
\frac{C_{\rm l}+\gamma_{\rm esc}- 
\sqrt{\gamma_{\rm esc}(\gamma_{\rm esc}+2C_{\rm l})}} 
{\sqrt{\gamma_{\rm esc}(\gamma_{\rm esc}+2C_{\rm l})}}.
\end{eqnarray}
In the limit of strong pumping this expression has the $\sqrt{C_{\rm l}}$
asymptotic behavior. The curve Eq.(\ref{33}) is represented in Fig.~4
by the solid line.


\section{Photovoltaic effect as pumping in phase space}


In this section we discuss the mechanism of charge transport by a
single pump at finite frequencies ({\it i.e.} the irreversible contribution to
the d.c.-current). We show the similarity with the  
mechanism of adiabatic charge pumping, discussed in \cite{Br}.
In the adiabatic approximation the system's motion is considered in a parameter
space. For finite frequencies the parameter space has to be
extended to phase space, which contains not only the perturbation
parameters but also their time derivatives.

According to Eq.~(\ref{10}), the transported charge for one period is
determined by 
\begin{eqnarray}
\displaystyle
Q & = & e\int_0^{T_{\rm p}}\!dt \int\! dt'\int d\tau\int\frac{d\e}{2\pi}
e^{i\e \tau} f(\tau) 
\label{12a}
\\
\displaystyle
& \times & 
{\rm Tr}
\left\{ {\cal S}\left(\e,\frac{t+t'}{2}+\frac{\tau}{4}\right)
{\cal
S}^{\dagger}\left(\e,\frac{t+t'}{2}-\frac{\tau}{4}\right)\Lambda\right\}.
\nonumber
\end{eqnarray}
We use the Wigner transform for the scattering matrix:
\begin{equation}
\label{12b}
{\cal S}(t,t')=\int {\cal S}(\e,(t+t')/2)e^{i\e (t-t')}\frac{d\e}{2\pi}.
\end{equation}

We consider charge pumping at high temperature ($T\gg \omega$). In
this case the integration over $\tau$ is limited by the inverse
temperature $1/T$. On the other hand, the scattering matrix ${\cal
S}(\e,t)$ in the Wigner
representation varies slowly with respect to its time argument $t$. 
This allows us to expand the scattering matrices in Eq.(\ref{12a}) 
to linear order in $\tau$. Using the unitarity of the scattering matrix 
we finally obtain
\begin{eqnarray}
\label{12c}
Q & = & e\int_0^{T_{\rm p}}dt\int\frac{d\e}{2\pi}\frac{1}{\cosh^2\e/2T} 
\\
\displaystyle
&\times &{\rm Tr}
\left\{ \Lambda \left( \frac{\partial {\cal S}\left(\e,t\right)}{\partial t}
{\cal S}^{\dagger}\left(\e,t\right)-
{\cal S}\left(\e,t\right)
\frac{\partial {\cal S}^{\dagger}\left(\e,t\right)}{\partial t}
\right) \right\}
\nonumber
\end{eqnarray} 
This equation was used by Brouwer in \cite{Br}. (See also
\cite{BTP}). The scattering matrix in the Wigner representation is a
function of the perturbation itself and its higher order derivatives with
respect to time. (See Appendix C.) In the adiabatic approximation the
derivatives are 
neglected as being small to higher orders in frequency. Beyond the adiabatic
approximation, we have to include the derivatives. 

We demonstrate that the analysis of Ref. \cite{Br} can be applied to
our case. We assume, that there is a single parameter $\varphi(t)$.
Then, following Brouwer, Ref. \cite{Br}, we introduce a vector field 
\begin{equation}
\label{12d}
P_i(\e,t)={\rm Im }\ {\rm Tr}\left\{
\Lambda \frac{\partial {\cal S}(\e,t)}{\partial X_i}{\cal S}^\dagger(\e,t)
\right\},
\end{equation}
where $X_i=d^i\varphi (t)/d t^i$ and $i$ is a
non--negative integer. 

In these notations Eq.(\ref{12c}) for the transported charge $Q$
is given by 
\begin{eqnarray}
\label{12c1}
Q & = & e\oint \int\frac{d\e}{2\pi}\frac{1}{\cosh^2\e/2T}
\sum\limits_{i=0}^{\infty} P_i(\e)dX_i.
\end{eqnarray}
The loop integral in the above equation can be rewritten as a surface
integral using Stoke's theorem. We develop our analysis for the
transported charge to the lowest order in frequency,
so that the scattering matrix depends only on 
$X_0=\varphi(t)$ and $X_1=\dot \varphi(t)$. According to Stoke's theorem
for this two dimensional space,
we obtain
\begin{eqnarray}
\label{area}
Q & = & e \int\frac{d\e}{2\pi}\frac{1}{\cosh^2\e/2T} \int d\varphi d
\dot \varphi \Pi(\e),
\\
\displaystyle
\Pi (\e)& = &
{\rm Im }\ {\rm Tr}\left\{
\Lambda \frac{\partial {\cal S}(\e)}{\partial \varphi}
\frac{\partial {\cal S}^\dagger(\e)}{\partial \dot\varphi}
\right\}.
\nonumber
\end{eqnarray}  
 
In appendix C we present a formal derivation of 
$\partial {\cal S}/\partial \dot \varphi$ from
the equation of motion, Eq.(\ref{14}), in terms of the Green's
functions of the dot to lowest order in $\omega/\gamma_{\rm esc}$.

Now we interpret results found in the previous section using
Eq.(\ref{area}). 
For weak pumping we keep the time dependent perturbation to 
lowest order to calculate the derivatives with respect to $\varphi$
and $\dot \varphi$. Then we consider $\Pi(\e)$ to be a constant and
the integral over $\varphi$ and $\dot\varphi$  gives the area of the contour 
in phase space. For a harmonic field $\varphi(t)=\cos\omega t$, the
contour is an ellipse with large semiaxis $\sqrt{C}$, small 
semiaxis $\omega \sqrt{C}$ and area $\pi\omega
C$. For the variance of the transmitted charge, we expect $\varQ 
\propto \omega^2 C^2/\gamma^4_{\rm esc}$, which is in agreement with
Eq.(\ref{30}) for $\omega\ll \gamma_{\rm esc}$.

In the limit of low frequency but strong pumping, we can apply
Eq.(\ref{area}) to understand Eq.(\ref{32}).
The power dependence [$C^{3/2}$]  is different from 
the adiabatic case [$C^{1/2}$]. 
The loop in phase plane is long
along the $\varphi$ axis but narrow in the $\dot \varphi$ 
direction because the frequency is small. [See Fig.~5(b).] 
The charge variation is determined by  
a sum of independent contributions from pieces of the contour
along the $\varphi$ axis. As can be seen from Fig.~5(b) the number of
the independent pieces is $N_{\rm ind}=\sqrt{C/\gamma_{\rm esc}}$.   
In the $\dot \varphi$ direction the system is correlated inside each piece
of the contour since all 
points along the $\dot\varphi $ direction are separated by a distance,
smaller than the correlation length.
The characteristic area $S_{\rm c}$ of each part is proportional to $\omega
\sqrt{C \gamma_{\rm esc}}$. 
The variance of the transported charge can thus be estimated as
\begin{equation}
\label{41a}
\varQ \propto e^2 N_{\rm ind} S^2_{\rm c} \propto e^2
\frac{\omega^2}{\gamma^2_{\rm esc}}
\left(\frac{C}{\gamma_{\rm esc}}\right)^{3/2}.
\end{equation}

When the amplitude of the field $C$ or the frequency $\omega$ increases 
further, so that $\omega^2
C_{\rm l} \geq \gamma^3_{\rm esc}$, this picture is no longer valid. The
trajectory does not have parts close to each other and each part gives
an independent contribution. The situation is similar to the case of
strong adiabatic pumping, as shown in Fig.~5(a) and discussed in
\cite{SAA}. The variance of the transported charge is proportional to
the total number of uncorrelated parts, so that $\varQ
\propto \sqrt{C}$, see Eq.~(\ref{33}).

  
\section{Low temperature}


The previous discussion of d.c.-current generation is quite
general. However, it does not take into account the heating
of electrons by an external field which becomes important 
at low temperature. 
In this regime, the electron distribution function in the dot
changes and acquires a width larger than the electron temperature 
in the leads.   

The new width $T_{\rm h}$ of the distribution function can be estimated 
from the following picture. An
electron has random transitions between different energy levels. The time
between consecutive transitions $t_{\rm tr}$ is determined by the Fermi golden
rule:
\begin{equation}
t^{-1}_{\rm tr}=\sum_m 2\pi |V_{nm}|^2\delta(\e_n-\e_m\pm \omega)\sim
\overline{ |V_{nm}|^2  }/\mls=\frac{C}{\pi}.
\end{equation}
The first equality sign follows from the Fermi golden rule, the second
sign represents an estimate of the characteristic value of the matrix
elements $\overline{|V_{nm}|^2}  $ and the density of states $1/\mls$,
the last equation is the definition of $C$, cf. Eq.(\ref{5}).
Since an electron stays in the dot for a time
$\tau_{\rm esc} = \gamma^{-1}_{\rm esc}=2\pi/N_{\rm ch}\mls$, 
it performs $N_{\rm tr}=1/t_{\rm tr}\gamma_{\rm esc}=C/\gamma_{\rm esc}$
transitions. Each transition changes the energy of the electron by
$\omega$. As in the random walk problem, the
displacement of electrons in energy space is $\propto
\omega\sqrt{C/\gamma_{\rm esc}}$.    

This analysis gives a new temperature scale $T_{\rm h}$:
\begin{equation}
\label{42a}
T_{\rm h}=\omega\sqrt{\frac{C}{\gamma_{\rm esc}}}.
\end{equation}
This scale has a meaning only for strong fields, $C\gg \gamma_{\rm esc}$,
so that the diffusion picture
in energy space is valid. Otherwise, electrons
experience few transitions with change of energy $\omega$.
Now we consider low temperatures, so that $T\gg T_{\rm h}$ is not valid.
We calculate the fluctuations of d.c. current for a system with a
single pump. As we know from Sec.~III, at high frequency the number of
pumps is not important and the result depends on their linear
combination.

Unlike the diagram, shown in Fig.~3, diagrams presented in Fig.~6 
have additional diffusons dressed on the distribution functions
$f(\tau)$. Collecting diagrams in Fig.~3 and Fig.~6 we obtain the following
expression for the variance of the pumped charge:
\begin{eqnarray}
\varQ & = & 4 e^2 CN_{\rm ch} g 
\int\!\!\!\int\limits_0^{T_p}\!\!dtdt'\int_0^{\infty}\!\!\! d\theta
\int_{-\theta}^{+\theta} 
\! \! d\tau \tilde F^2(\tau)
\nonumber
\\
\label{42}
&\times &
\D\left(\frac{t+t'}{2}+\theta ,\frac{t+t'}{2}+\tau, t-t' \right)
\\
&\times &
\D\left(\frac{t+t'}{2}+\theta ,\frac{t+t'}{2}-\tau, t-t' \right)
\nonumber
\\
&\times &
\int_0^{+\infty}\!\!\!d\xi  \D(t,t-\xi,2 \tau)
\int_0^{+\infty}\!\!\!d\xi' \D(t',t'-\xi',2 \tau)
\nonumber
\\
&\times &
\left\{
 2C \sin^2\o(t-\xi) \sin^2\o(t'-\xi')\sin^2\o\tau
\right.
\nonumber
\\
& &
+\left.  N_{\rm ch} \sin\o t \sin \o t'\right\}
\nonumber
\end{eqnarray}
where $g$ is the dimensionless conductance of the dot (see Eq.(\ref{g})) and  
\begin{equation}
\label{42F}
\tilde F(\tau)=\frac{T\sin \omega \tau}{\sinh 2\pi T\tau}.
\end{equation} 

At high temperature $T\gg \o$ Eq.(\ref{42}) reduces to Eq.(\ref{22}).
Indeed, all three diagrams in Fig.~6 are smaller than the diagram in
Fig.~3 at least by one factor $\o/T$.

Now we discuss the limit of high frequency $\o\gg{\rm max}\{\gamma_{\rm esc},
C\}$. This inequality allows us to perform integration over $\xi$ and
$\xi'$ in Eq.(\ref{42}). We can replace $\sin^2\o(t-\xi)$ by $1/2$ and use
the approximation
\begin{equation}
\label{43}
\int_0^\infty \D(t,t-\xi,2\tau)d\xi\approx
\frac{1}{N_{\rm ch}+2C\sin^2\omega\tau}.
\end{equation}
In this limit the product of the diffusons in the second and third
lines of Eq.(\ref{42}) does not depend on $\tau$. 

Energy of an electron in the dot changes due to the external field
resulting in the redistribution of the electrons in the energy space.
The new distribution function becomes wider than that of
electrons in the leads at temperature $T$. Consequently,
the Fourier transform of the electron distribution function becomes
narrower. The right hand side of Eq.(\ref{43}) represent the effect of
heating. This function appears in the integral over $\tau$ in
Eq.(\ref{42}) along with the function $F(\tau)$, defined by
Eq.(\ref{223}). At sufficiently low temperature the convergence of the
integral over $\tau$ is determined by the 'heating factors',
Eq.(\ref{43}) rather than by $F(\tau)$. We note, that the shape of the new 
distribution function is not a Fermi function with a higher temperature.
Instead, its Fourier transform has the form of the right hand side of 
Eq.~(\ref{43}).

\begin{figure}
\centerline{\psfig{figure=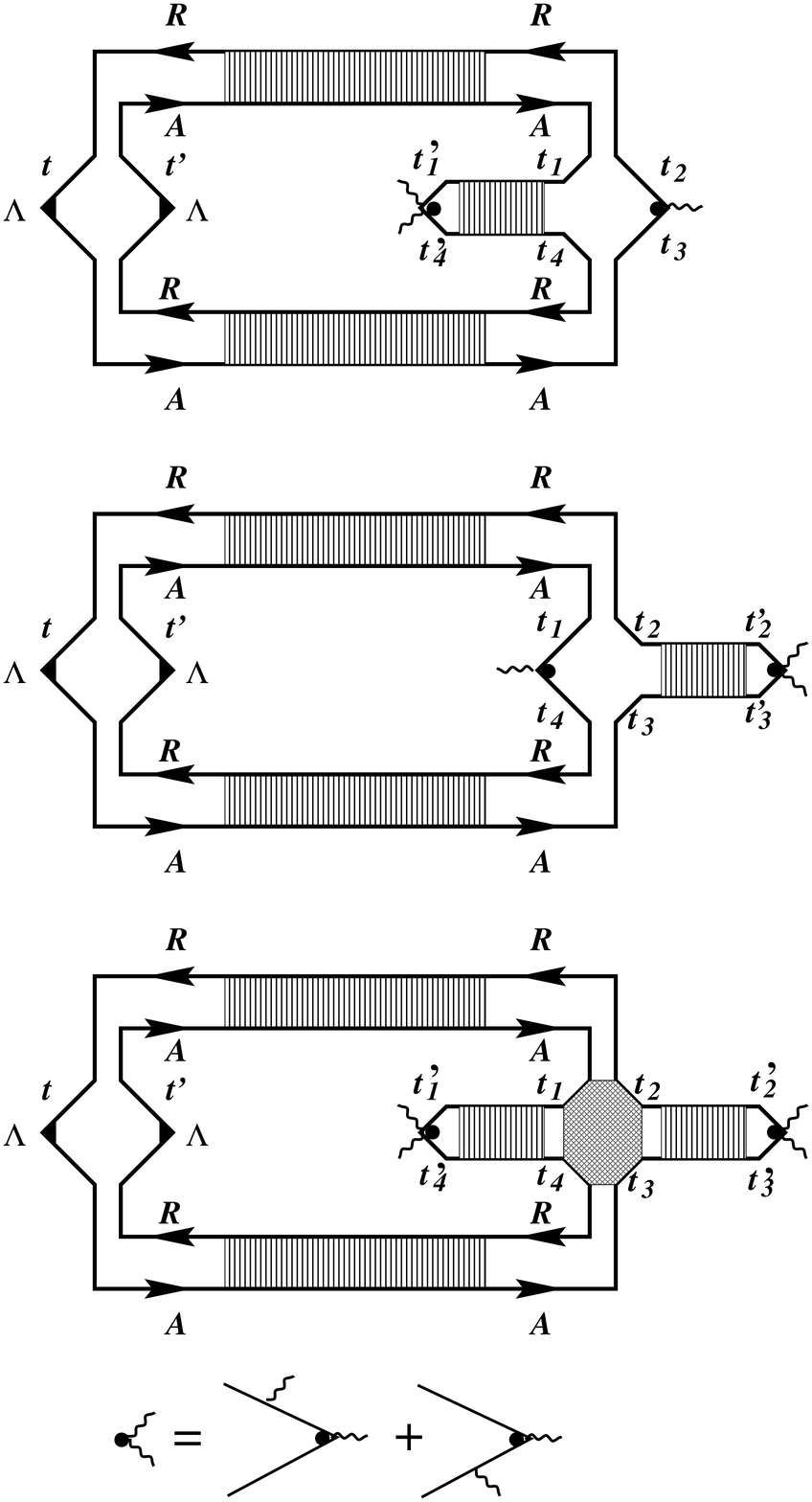,width=8cm}}
\label{f6}
\narrowtext{
\caption{Diagrams, which contribute to the d.c.-current 
at low temperature limit. (We do not show diagrams, which can be obtained
from the above by omitting the upper or lower diffusons.) The last
diagram contains the Hikami box, which is presented in Fig.~7.}}
\end{figure}

To be more specific, we consider the strong pumping limit $C\gg
\gamma_{\rm esc}$, when
the electron distribution function is determined by the new
scale $T_{\rm h}$, see Eq.(\ref{42a}). We find
\begin{equation}
\label{45}
\varQ  =\frac{3}{16}e^2 g \frac{\mls}{\omega}.
\end{equation}
The same parameter dependence can be found from Eq.(\ref{33}) 
replacing temperature $T$ by the new energy scale $T_{\rm h}$.

Equation~(\ref{42}) has two terms. One term contains factor
$\sin\o t\sin \o t'$. This term survives the high temperature limit,
see Sec.~III. Nonetheless at low temperature the heating modifies the
results of Sec.~III, so that the temperature dependence saturates at
the characteristic temperature scale $T_{\rm h}$. 

The second term was completely neglected in the previous sections. The
origin of this term is similar to that of the thermoelectric effect in
a conductor out of thermodynamic equilibrium. Although electrons
are in equilibrium in the leads, the heating changes their distribution
function in the dot, producing a non-equilibrium
distribution. Then non-equilibrium electrons escape from the dot. The
direction of each escape is determined by the realization of the
dot. An unbalance between electrons escaping through the left or right
leads gives current. ($\sin^2\omega \tau$ term in Eq.~(\ref{42})
reflects the electron--hole asymmetry, necessary for
thermoelectric  effects.)

\begin{figure}
\centerline{\psfig{figure=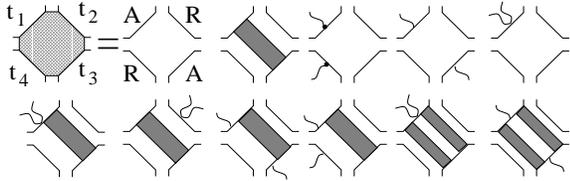,width=7.5cm}}
\label{f7}
\narrowtext{
\caption{
The Hikami box, introduced in Fig.~6, can be obtain from these
diagrams as well from their rotations. The grey rectangulars represent
averages of the type $\overline{\R \R}$ and $\overline{\A\A}$.}}
\end{figure}


\section{Conclusions}


We studied the d.c.-current through the quantum dot generated  
{\it e.g.} by time-dependent distortions of the dot shape. 
This d.c.-current is fluctuating from sample to sample and we found
the second moment of its distribution. Unlike the previous works
\cite{Br,SAA} on the adiabatic pumping we treated the system for a
broad range of external frequencies thus providing  a bridge between
adiabatic pumping and photovoltaic effects in microjunctions of
Ref.~\cite{FK}. 

The adiabatic approximation is not valid when the frequency of the
perturbation $\omega$ is comparable with the escape rate from the dot
$\gamma_{\rm esc}$. Beyond the adiabatic approximation the
d.c.-current consists of reversible and irreversible parts. The
reversible contribution was studied in the adiabatic
regime. This contribution is determined by an integral over the
contour in parameter space. On the other hand the irreversible
contribution to the d.c.-current is determined by an integral over this
contour in phase space. 

A crossover from the bilinear ($C^2$) to
$\sqrt{C}$ regime was found in \cite{SAA}. The crossover happens when
the system makes a large loop during one period of the external field,
so that it is uncorrelated at different points of the loop, see 
Fig.~5(a). We showed, that this crossover is universal and 
happens at arbitrary frequency. This result is
consistent with the representation of the d.c.-current as an integral
along the contour in the phase space. 
An intermediate regime exists for a single pump at low frequency and
moderate amplitude of the external field. This regime is described by
a $C^{3/2}$ dependence of the variance of the d.c.-current on 
the field amplitude.

We also considered a wide temperature range. At high temperature the
variance of the current decreases as $T^{-1}$. 
We found that at low temperature heating becomes important and it
introduces a characteristic temperature $T_{\rm h}$, see
Eq.~(\ref{42a}), below which the temperature dependence of the
d.c.-current saturates. The result of heating on the d.c.-current is
twofold. The first effect diminishes the d.c.-current by broadening of the
electron distribution function. The second effect produces the
a thermoelectric field, and is a non-equilibrium effect. This effect is
related to the electron-hole asymmetry in the dot.
Our results are different from the observed experimentally temperature
dependence, see also Ref.~\cite{SAA}.

Finally, the photovoltaic effect is not symmetric with respect to
inversion of magnetic field, similarly to adiabatic pumping case, see
Ref.~\cite{SAA}.

We are thankful to P.~W.~Brouwer for useful discussions.
The work was supported by Cornell Center for Materials Research under
NSF grant No. DMR-9632275 (M.G.V. and V.A.) and Packard Fellowship 
(I.L.A.). 
\appendix
\section{}

We define the wave function of electrons in channel $\alpha$ moving
towards the dots by $\psi_{\alpha}(x,t)$ with $x<0$, where $|x|$
determines the distance from the dot boundary, see Fig.(8). Then
$\psi_{\alpha}(x,t)$ for $x>0$ represents the outcoming electrons. The
boundary $x=0$ is described by a superposition of the incoming and
outcoming  electron states and we denote it by $\psi_{\alpha}(0,t)$.
The wave function of electrons in state $i$ is denoted by $\psi_i(t)$.

\begin{figure}
\centerline{\psfig{figure=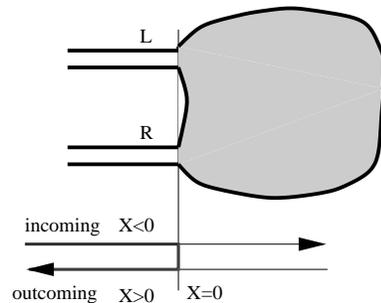,width=5cm}}
\label{f8}
\narrowtext{
\caption{
Correspondence between the sign of $x$ and the direction of motion of
electrons with respect to the dot. }}
\end{figure}

We introduce the Keldysh Green's functions 
\begin{eqnarray}
\label{a1}
\hat {\cal G}_{\alpha\beta}(t,t',x,x') &=&\\
& &\left( 
\begin{array}{cc}
{\cal G}^{(R)}_{\alpha\beta}(t,t',x,x') & 
{\cal G}^{(K)}_{\alpha\beta}(t,t',x,x')\\
0 & {\cal G}^{(A)}_{\alpha\beta}(t,t',x,x')
\end{array}
\right),
\nonumber
\\
\label{a2}
\hat {\cal G}_{i\alpha}(t,t',x') & =&
\left(
\begin{array}{cc}
{\cal G}^{(R)}_{i\alpha}(t,t',x') & 
{\cal G}^{(K)}_{i\alpha}(t,t',x')\\
0 & {\cal G}^{(A)}_{i\alpha}(t,t',x')
\end{array}
\right),
\end{eqnarray}
which are defined in terms of 
\begin{eqnarray}
{\cal G}^{(R)}_{\alpha\beta}(t,t',x,x') & = &
-i\Theta(t-t')\langle [\psi_{\alpha}(x,t),
\psi^\dag_{\beta}(x',t')]_+\rangle, 
\nonumber
\\
{\cal G}^{(A)}_{\alpha\beta}(t,t',x,x') & = &
i\Theta(t'-t)\langle [\psi_{\alpha}(x,t),
\psi^\dag_{\beta}(x',t')]_+\rangle,
\nonumber
\\
{\cal G}^{(K)}_{\alpha\beta}(t,t',x,x') & = &
-i\langle [\psi_{\alpha}(x,t),
\psi^\dag_{\beta}(x',t')]_-\rangle,
\nonumber
\end{eqnarray} 
where $[\cdot,\cdot]_{\pm}$ denote commutator and anticommutator
respectively. The similar expressions can be written down for 
$\hat {\cal G}_{i\alpha}(t,t',x')$ Green's function, with
$\psi_{\alpha}(x,t)$ replaced by $\psi_i(t)$.

We assume that electrons do not interact in the reservoirs and 
the Green's function of the incoming electrons ($x;\ x'<0$) 
is given by the Keldysh structure:
\begin{eqnarray}
\label{a3}
& &
\displaystyle
{\cal G}_{\alpha\beta}(t,t',x,x')\\
\displaystyle
\nonumber
&=&\left(
\begin{array}{cc}
\R_{\alpha\beta}(t-t',x-x') & \K_{\alpha\beta}(t-t',x-x')\\
0 & \A_{\alpha\beta}(t-t',x-x')
\end{array}
\right),
\end{eqnarray}
where
\begin{eqnarray}
\label{a4}
\displaystyle
\R_{\alpha\beta}(t,x)&=&i\Theta(t)\ \delta_{\alpha\beta} \delta
\left( v_F t-x \right),\\
\label{a5}
\displaystyle
\A_{\alpha\beta}(t,x)&=&-i\Theta(-t)\ \delta_{\alpha\beta} \delta
\left( v_Ft-x\right),\\
\label{a6}
\displaystyle
\K_{\alpha\beta}(\e,x)&=&\tilde f_\alpha(\e)\left(
\R_{\alpha\beta}(\e,x)-\A_{\alpha\beta}(\e,x)
\right),
\end{eqnarray}
where $\tilde f(\e)$ is the distribution function of electrons in the
channel
$\alpha$. In equilibrium with temperature $T$, 
\begin{equation}
\tilde f_\alpha(\e)
=\tanh\frac{\e-\delta \mu_\alpha}{2T},
\end{equation}
where $\delta \mu_\alpha$ represent relative change in chemical
potential for different leads.  

The equations of motion for the Green's functions defined by
Eqs.~(\ref{a1}) and (\ref{a2}) have the form:
\begin{eqnarray}
\label{a7}
& & i\left[\frac{\partial}{\partial t}-v_F\frac{\partial}{\partial
x}\right]
\hat {\cal G}_{\alpha\beta}(t,t',x,x') \ \ \ \ \ \
\\
\nonumber
& &
\ \ \ \
=\delta(x)W_{\alpha i}\hat {\cal G}_{i\beta}(t,t',x')
+\delta(t-t')\delta(x-x')\hat 1,
\\
\label{a8}
& &
\left[i\frac{\partial}{\partial t}-H_{ij}(t)\right]\hat {\cal
G}_{j\alpha}(t,t',x')
\\
\nonumber
& &
\ \ \ \
= W^\dag_{i\beta}\hat {\cal G}_{\beta\alpha}(t,t',0,x').
\end{eqnarray}

We notice that due to causality, $\A_{\alpha\beta}(t,t',0,x')\equiv 0$
for $x'<0$. This observation
significantly simplifies further calculations. Indeed, we can
represent the Keldysh component of the Green's function in the left
hand side of
Eq.(\ref{a8}) in the form
\begin{eqnarray}
\label{a9}
{\cal G}^{(K)}_{i\alpha}(t,t',x')& = &\int
dt_1\left[\frac{1}{i\partial/\partial t -\hat
H(t)}\right]_{ij}(t,t_1)\\
&\times &
W_{j\beta}^\dag{\cal G}_{\alpha\beta}(t_1,t',0,x'),
\nonumber
\end{eqnarray}
The corresponding advance component is zero. Here
$1/(i\partial/\partial t -\hat H(t))$ is the retarded component of
the electron Green's function in the dot. This definition is different
from that given in the main part of the paper, see Eq.(\ref{11}).
The latter will appear naturally in the end of this section. The
additional term $\sim W^\dag W$ takes into account the escape from the
dot through the leads.

The next step is to represent Eq.(\ref{a7}) in the form
\begin{eqnarray}
\label{a10}
& &
{\cal G}^{(K)}_{\alpha\beta}(t,t',x,x')=\K_{\alpha\beta}(t-t',x-x')
\\
& &
+ \int dt_1dt_2 \R_{\alpha\gamma}(t-t_1,x)
\left[W\frac{1}{i\partial/\partial t -\hat
H(t)}W^\dag \right]_{\gamma\delta}\!\!\! (t_1,t_2)
\nonumber
\\
& &
\ \ \ \ \ \  \
\times
{\cal G}^{(K)}_{\delta\beta}(t_2,t',0,x')
\nonumber
\end{eqnarray}

In the above equation we consider $x=0$. Using
$\R_{\alpha\beta}(t-t',0)$ from Eq.(\ref{a4}) we find
\begin{eqnarray}
{\cal G}^{(K)}_{\alpha\beta}(t,t',0,x')&=&\!
\int\!\! dt_1
\left[1-\hat W \frac{i\pi\nu}{i\partial/\partial t -\hat
H(t)}\hat W^\dag \right]^{-1}_{\alpha\delta}\!\!\!\! (t,t_1)
\nonumber
\\
\label{a11}
&\times&  G^{(K)}_{\delta\beta}(t_1,t',0,x'), \ \ x'<0.
\end{eqnarray}

Substituting this expression to Eq.(\ref{a10}) and taking $x=+|\delta|\to
0$, we obtain for $x'<0$:
\be
\label{a12}
{\cal G}^{(K)}_{\alpha\beta}(t,t',+|\delta|,x')=\int\!\! dt_1{
\cal S}_{\alpha\gamma}(t,t_1)\K_{\gamma\beta}(t_1-t',-x'),
\ee
where the scattering matrix ${\cal S}_{\alpha \beta}(t,t')$ is given
by Eq.(\ref{13}).

Equation (\ref{a12}) is valid for $x'<0$. We have to repeat the procedure
described above to calculate the electron Green's function in the
leads for $x'>0$.
Since the equations which determine evolution of the Green's function
from $x'<0$ to $x'>0$ are conjugated to those for $x$, we conclude,
that
\begin{eqnarray}
& &
\label{a13}
{\cal G}^{(K)}_{\alpha\beta}(t,t',+|\delta|,+|\delta|)
\\
& &\ \ \ \ \ \nonumber =
\int\!\!\int\!\! dt_1dt_2
{\cal S}_{\alpha\gamma}(t,t_1)\K_{\gamma\delta}(t_1-t_2,0)
{\cal S}^\dag_{\delta\beta}(t_2,t')
\end{eqnarray}

The currents in the left and right leads are given by
\begin{eqnarray}
& &
I_{\rm l}(t)=ev_{\rm F}
\label{a14}
\\
& & \ \nonumber \times
\sum\limits_{\alpha=1}^{N_{\rm l}}\!\!\left(
{\cal G}^{(K)}_{\alpha\alpha}(t,t,+|\delta|,+|\delta|)-
{\cal G}^{(K)}_{\alpha\alpha}(t,t,-|\delta|,-|\delta|)
\right)\nonumber
\\
& &
I_{\rm r}(t)=ev_{\rm F}
\label{a15}
\\
& & \
\times
\sum\limits_{\alpha=N_{\rm l}+1}^{N_{\rm ch}}\!\!\left(
{\cal G}^{(K)}_{\alpha\alpha}(t,t,+|\delta|,+|\delta|)-
{\cal G}^{(K)}_{\alpha\alpha}(t,t,-|\delta|,-|\delta|)
\right),
\nonumber
\end{eqnarray}
where $\delta\to 0$. This limit is just a reminder that 
${\cal G}^{(K)}_{\alpha\alpha}(t,t,-|\delta|,-|\delta|)$ is taken for
incoming electrons and is given by Eq.(\ref{a6}). Consequently, 
\begin{eqnarray}
\label{a13'}
{\cal G}^{(K)}_{\alpha\alpha}(t,t,-|\delta|,-|\delta|)&=&
f(+i0),\\
f(t)=\int_{-\infty}^{+\infty} e^{i\omega t}\tilde
f(\omega) \frac{d\omega}{2\pi}.
\nonumber
\end{eqnarray}

Since the charge is conserved, $I_{\rm l}(t)=-I_{\rm r}(t)$. We rewrite the
current through the dot as
\be
\label{a16}
I(t)=\frac{N_{\rm r}I_{\rm l}(t)-N_{\rm l}I_{\rm r}(t)}{N_{\rm
ch}}=I_{\rm l}(t)=-I_{\rm r}(t). 
\ee
Substituting Eqs.~(\ref{a13}) and (\ref{a13'}) into Eqs.~(\ref{a14}) 
and (\ref{a15}) and using Eq.(\ref{a16}) we obtain Eq.(\ref{10}).

\section{}
In this appendix we derive Eq.(\ref{19}) from the general Eq.~(\ref{12}).
The only assumption we are using here is that the distribution
function of electrons is the same in all channels, {\it i.e.}
$f_\alpha(t)\equiv f(t)$.
Substituting the explicit form of the scattering matrix from
Eq.~(\ref{13}), we obtain
\begin{eqnarray}
I(t)&=&2\pi i \nu e \left\{
\int\! dt_1 f(t-t_1) \Lambda_{\alpha\beta}\left[
\hat W^\dagger \hat G^{(A)}(t_1,t)\hat W
\right]_{\beta\alpha}\right.
\nonumber
\\
&-&\int\! dt_1 f(t_1-t) \Lambda_{\alpha\beta}\left[
\hat W^\dagger \hat G^{(R)}(t,t_1)\hat W
\right]_{\beta\alpha}
\label{b1}
\\
&-&2\pi i \nu \int\!\!\int\!\! dt_1 dt_2 f(t_1-t_2)
\Lambda_{\alpha\beta}
\nonumber
\\
&\times &\left. \left[
\hat W^\dag \hat G^{(R)}(t,t_1)\hat W
\hat W^\dag \hat G^{(A)}(t_2,t)\hat W
\right]_{\beta\alpha}\right\},
\nonumber
\end{eqnarray}
where the diagonal matrix $\hat \Lambda $ is defined in
Eq.~(\ref{13}), and $\hat G(t,t')$ denotes the Green's function in the dot.

Now we use the equation of motion of the retarded and advanced Green's
functions, Eq.~(\ref{14}), to simplify the right hand side of
Eq.~(\ref{b1}). For this purpose, we pre--multiply the equation for the
advanced component $\hat G^{(A)}(t_2,t')$ by the retarded component  
$\hat G^{(A)}(t,t_1)$, then we
post--multiply the transposed equation for the retarded component  
$\hat G^{(A)}(t,t_1)$ by $\hat G^{(A)}(t_2,t')$. 
Subtracting from the second equation the first one, we obtain
\begin{eqnarray}
& &
2\pi i \nu 
\hat G^{(R)}(t,t_1)\hat W \hat W^\dag \hat G^{(A)}(t_2,t')
\label{b2}\\
& = &
i\left[\frac{\partial}{\partial t_1}+
\frac{\partial}{\partial t_2}\right]\hat G^{(R)}(t,t_1)
\hat G^{(A)}(t_2,t')
\nonumber
\\
& + &
\hat G^{(R)}(t,t_1)\left[\hat H(t_1)-\hat H(t_2)\right]
\hat G^{(A)}(t_2,t')
\nonumber
\\
&  - &
\left[\hat G^{(R)}(t,t_1)\delta(t_2-t')-\delta(t-t_1)\hat
G^{(A)}(t_2,t')\right]
\nonumber
\end{eqnarray}

Substituting this equation to Eq.~(\ref{b1}) we obtain Eq.(\ref{19}).

\section{}

The equation for the Green's function in the Wigner
representation for a time--dependent Hamiltonian is 
\begin{eqnarray}
\label{c1}
\displaystyle  
&&  2  
\e \hat G(\e,t)- \left[ H_0-i\pi \nu \hat W\hat W^\dagger,\hat G(\e,t)
\right]_+
\\
& + &
\sum_{k=0}^{\infty} \frac{\varphi^{(k)}(t)}{(2i)^k k!}
\left\{ \hat V 
\frac{\partial^k \hat G(\e,t)}{\partial \e^k}  
+(-1)^k \frac{\partial^k\hat G(\e,t)}{\partial \e^k }\hat V\right\} = 2.
\nonumber
\end{eqnarray}
Here $\hat G(\e, T)$ is the Green's function in the Wigner variables $\e$
and $t$. We represented $\hat H(t)$ in the form $\hat H(t)=\hat
H_0+\hat V \varphi(t)$.

In the adiabatic limit only the $k=0$ is taken into account. This
approximation is crucial in the case of a single pump. By
appropriate choice of the beginning of the cycle the pump moves for the second
half of the cycle along the same trajectory as for the
first half, but in the opposite direction. As a consequence, the total
transported charge $Q$, Eq.(\ref{12c}), vanishes in the adiabatic
approximation. To remove this symmetry, we can add another pump
oscillating with the same frequency, but with different phase shift.
Also, the higher order terms in Eq.(\ref{c1}) break this symmetry.   

We consider contribution to the lowest order in frequency to the
transported charge $Q$. For this purpose,  we neglect all terms with
$k\geq 2$, keep the $k=1$ term to the first order and include all
orders in $k=0$ term. The solution is 
\begin{eqnarray}
\label{c2}
\hat G_0(\e,t) & = & \frac{1}{\e-\hat H_0-\hat
V\varphi(t)+i\pi\nu\hat W\hat W^\dagger}, \\
\hat G_1(\e,t) &=&i\frac{\dot \varphi(t)}{2}\left(\hat G_0(\e,t)\hat V
\hat G^2_0(\e,t)- 
\hat G^2_0(\e,t)\hat V \hat G_0(\e,t)\right).
\nonumber
\end{eqnarray}

To the lowest order in $\dot \varphi(t)$, the scattering 
matrix $\hat {\cal S}$ has the form
\begin{equation}
\label{c3}
\hat {\cal S}_1(\e,t)=\hat{\cal S}_0(\e,\varphi(t))+ 
\dot \varphi(t)\hat{\cal A}(\varphi(t)),
\end{equation}
where 
\begin{equation}
\label{c4}
\hat{\cal S}_0(\e,\varphi(t))  =  1-2\pi\nu \hat W^\dagger \hat
G_0(\e,t)\hat W,
\end{equation}
and 
\begin{eqnarray}
\label{c8}
\hat {\cal A}(\e,t) & \equiv & 
\frac{\partial \hat {\cal S}}{\partial \dot\varphi}
\\
\nonumber
& = &
\frac{i}{2}\left(\hat G_0(\e,t)\hat V
\hat G^2_0(\e,t)- 
\hat G^2_0(\e,t)\hat V \hat G_0(\e,t)\right).
\end{eqnarray}

\end{multicols}

\end{document}